\documentclass[review,number,sort&compress]{elsarticle}

\journal{Nuclear Instruments and Methods A}

\begin{document}

\begin{frontmatter}

\title{Tracking algorithms for the active target MAYA}

\author[label1,label2]{T. Roger\corref{cor1}}
\ead{thomas.roger@fys.kuleuven.be}
\author[label3]{M. Caama\~no}
\author[label4]{C.E. Demonchy}
\author[label5]{W. Mittig}
\author[label1]{H. Savajols}
\author[label6]{I. Tanihata}
\address[label1]{GANIL, Bd Henri Becquerel, BP 55027, F-14076 Caen Cedex 05, France}
\address[label2]{Instituut voor Kern- en Stralingsfysica, K.U. Leuven, Celestijnenlaan 200D, B-3001 Leuven, Belgium}
\address[label3]{Universidade de Santiago de Compostela, E-15786 Santiago,, Spain}
\address[label4]{CENBG-Université Bordeaux 1-UMR 5797 CNRS/IN2P3, Chemin du Solarium, BP 120, F-33175 Gradignan Cedex, France}
\address[label5]{NSCL, MSU, East Lansing, Michigan 48824, USA}
\address[label6]{RCNP, Osaka University, Mihogaoka, Ibaraki, Osaka 567 0047, Japan}
\cortext[cor1]{Corresponding author}

\begin{abstract}
The MAYA detector is a Time-Charge Projection Chamber based on the concept of active target. These type of devices use a part of the detection system, the filling gas in this case, in the role of reaction target. The MAYA detector performs three-dimensional tracking, in order to determine physical observables of the reactions occurring inside the detector. The reconstruction algorithms of the tracking use the information from a two-dimensional projection on the segmented cathode, and, in general, they need to be adapted for the different experimental settings of the detector. This work presents some of the most relevant solutions developed for the MAYA detector.
\end{abstract}

\end{frontmatter}

\section{Introduction}
\label{Intro}

Nowadays, the development of new radioactive beams allows nuclear physics to explore more exotic regions of the nuclear chart, revealing more new properties as they become experimentally available. The access to these regions usually involve exotic with low intensity and reactions with small cross-sections that force to improve detection and analysis techniques. To overcome these difficulties, experimental setups focus on different solutions, such as high efficiency and signal-to-noise discrimination, and the use of thick targets. Active target detectors, i.e, detection devices that use part of their systems as reaction target, proved to match these needs: since the detection is done inside the target, detection efficiency and effective target thickness are increased without losing resolution due to reaction point indetermination.

The concept of active target, developed more than fifty years ago in high-energy physics uses, is being progressively adapted for its application in nuclear physics. The archetype of active targets in the domain of secondary beams is the detector IKAR \cite{DOB83}, used at GSI (Germany) to study elastic scattering of exotic beams at relativistic energies. Another example is the MSTPC detector \cite{MIZ99} designed at RIKEN (Japan) to study fusion and astrophysical nuclear reactions in low-energy regions. Presently, new designs are mostly based on gas-filled devices where the gas constitutes both the target and the detection medium. Among these, MAYA \cite{DEM02,DEM07}, developed and built at GANIL, is designed to explore very low energy domains not accessible with the use of solid or liquid targets. The MAYA detector applies the concept of Charge and Time Projection to perform a full three-dimensional reconstruction of the detected reaction with the charge collected in a segmented cathode and its associated drift time.

Most of the active targets in development use a similar configuration, with the tracking performed on a segmented layer. Therefore, some of the problems and solutions that appear in the reconstruction process are common to these detectors. In the case of MAYA, the tracking process needs to be adapted to the experimental configurations used to study different reactions, producing a collection of reconstruction protocols to extract the relevant observables. Among these, the angle, reaction vertex, and stopping points need specific formulas to be determined. Here, the most significant of these algorithms are reviewed. 

\section{The MAYA Detector}

\begin{figure}
\begin{center}
\includegraphics[width=12cm]{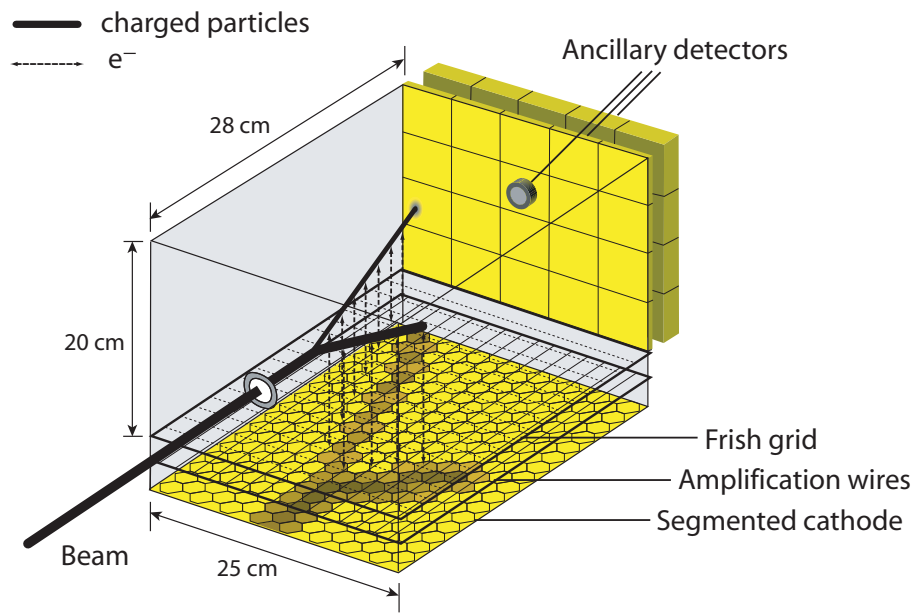}
\caption{\label{Fig.1} (Color online) The picture shows a schematic rendition of the MAYA active-target. A beam projectile enters the detector volume where it reacts with a nucleus in the gas. The particles involved in the reaction may produce enough ionization to induce a pattern in the segmented cathode, after traversing a Frisch grid and a plane of amplification wires. A set of ancillary detectors is used in the exit side of the detector.}
\end{center}
\end{figure}

Figure \ref{Fig.1} shows a typical MAYA setup. Two main zones can be identified within the detector: an active volume of 28$\times$26$\times$20 cm$^3$ where the reaction takes place, and the amplification area where detection and readout occur. The amplification zone consists of a Frisch grid, an anode wire plane below, and a segmented cathode in the lower part. The cathode is segmented into 33$\times$32 hexagonal pads, each of which measures 5 mm per side, arranged in rows parallel to the anode wires.

In general, the detection occurs when the beam particles and the reaction products ionize the filling gas along their paths. The electrons released in the ionization process drift toward the amplification area where they are accelerated in the vicinity of the wires, inducing mirror charges on the corresponding pads, which are measured and coded individually. Typically, the image charge from one avalanche will spread over several pads and the resulting distributions are used to obtain a two dimensional projection of the tracks of charged particles.

Measurements of the drift time of the ionizing electrons up to the amplification wires allow to calculate the vertical position. This information is combined with the reconstruction of trajectories projected on the cathode plane to perform a complete 3-dimensional tracking of the reaction products that lose enough energy to be detected. Ancillary detectors, such as cesium iodide crystals \cite{CAA07}, silicon \cite{MON08,ROG08,TAN08}, or diamond detectors are usually placed at the back, corresponding to forward angles in order to detect particles that do not stop inside the gas volume. Also, stoppers are employed for non-reacting beam particles that do not stop in the filling gas. Other modifications include beam-shielding \cite{MON08} and a modified drift chamber placed before the ancillary detectors.

The filling gas is chosen according to the reaction of interest. So far, MAYA was operated and tested with $^{2}$H$_{2}$ or $^{4}$He, either pure or mixed with standard detection gases such as methyl-propane C$_{4}$H$_{10}$ or CF$_4$, at pressures between \mbox{20 mbar} and \mbox{1 atm}.

The trajectory reconstruction from the sampled positions in the segmented cathode requires different algorithms that may vary from one configuration to the other. The tracking techniques extract information such as projected angles of trajectories, the position of the reaction vertex, and the determination of the stopping points, which are necessary to determined the range of the particles inside the gas.

\section{Two-Dimensional Charge Distributions}
\label{MAYA} 
\begin{figure}
\begin{center}
\includegraphics[width=12cm]{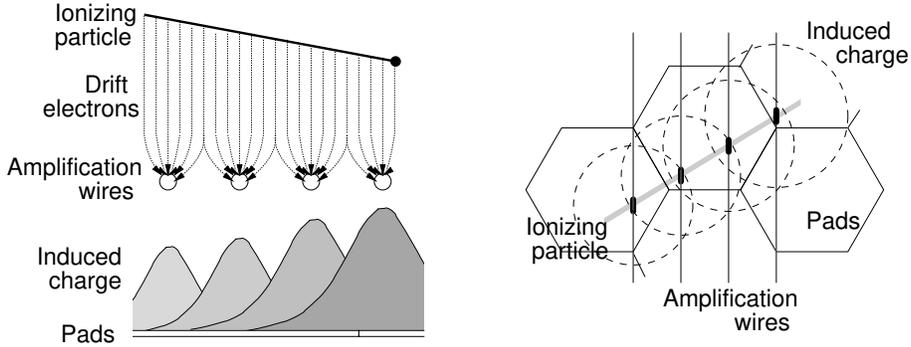}
\caption{\label{Fig.2} The processes involved in the formation of the two-dimensional pattern in the cathode plane are schematically summarized in the picture. Left drawing is a vertical scheme of the ionization, digitalization, and charge induction. Right figure shows the same processes in the horizontal plane.}
\end{center}
\end{figure}

\begin{figure}
\begin{center}
\includegraphics[width=12cm]{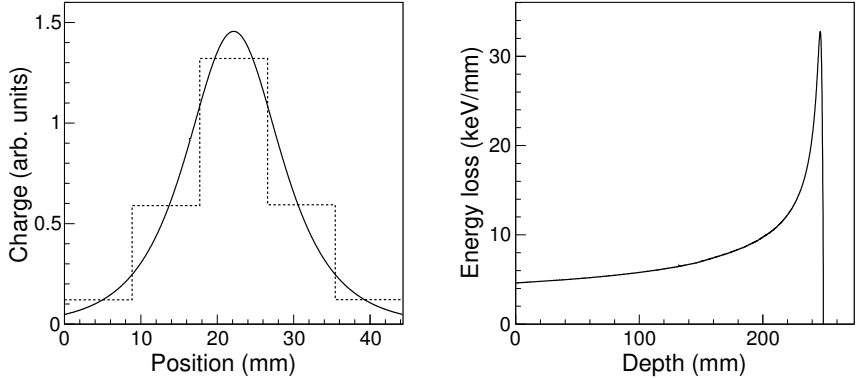}
\caption{\label{Fig.3} Left panel: Simulated charge distribution created by a point source along one dimension (solid line), integrated over pads (dashed line). The distribution corresponds to $L$=10 cm in Eq. \ref{ind}. Right panel: Energy-loss curve of a \mbox{2 MeV} proton in \mbox{1 atm} of isobutane generated by the TRIM code.}
\end{center}
\end{figure}

The two-dimensional projection of the particle trajectories on the cathode plane can be described as the convolution of different processes: the ionization path is digitized perpendicularly to the beam direction as the released electrons are attracted to the amplification wires; the amplification process induces a mirror charge on the pads below the wires that can be described as produced by multiple point-like sources; these are weighted by the energy-loss of the particles; and finally the resulting induced charge is integrated in the hexagonal-shape of each pad. Fig. \ref{Fig.2} summarizes these processes. These steps are reproduced in a simulation of the entire process, providing realistic patterns where different algorithms can be tested to reconstruct the original tracks. The simulation code generates two-dimensional patterns by reproducing the different processes:

- The energy-loss along the particle trajectory for different ionizing particles, energies, and gas compositions and pressures is obtained from Monte Carlo simulations using the TRIM code \cite{SRIM}. A typical energy-loss profile of a \mbox{2 MeV} proton in \mbox{1 atm} of isobutane is presented in Fig. \ref{Fig.3}. The calculated energy-loss profiles are projected (digitized) along the wires to determine the total charge induced, $Q$ in Eq. \ref{ind}, by each point-like source along the trajectory. The straggling of the electrons inside the gas is not yet included. For E/P $>$ 0.8 V.cm$^{-1}$.Torr$^{-1}$, it has been estimated to be less than 1 mm.

- The induction from a point-like source can be expressed as an exact electrostatic formula, as it is shown in Ref. \cite{END81}: 

\begin{equation} 
\sigma(x,y)=\frac{-Q}{2\pi}\sum_{n=0}^{\infty}\frac{(-1)^{n}(2n+1)L}{[(2n+1)^{2}L^{2}+x^{2}+y^{2}]^{3/2}} 
\label{ind}
\end{equation}

where $Q$ is the total charge, $L$ is the distance between the point-like source and the observation plane, and $x,y$ is the position with respect to the source. A typical charge distribution created by a point-like source is shown in Fig. \ref{Fig.3}. 

- Finally, the charge-induced distributions from all point-like sources is integrated on the surface of each pad to obtain the charge measured. 

The reconstruction algorithms are tested on sets of data that reproduce different experimental conditions in MAYA. These are classified depending on the particles detected on the cathode as:

- Single-track setups tag those where beam particles do not produce any charge pattern on the cathode plane, either because its energy-loss is too small (see for example \cite{CAA07}), or because electrons created by the incident ions are stopped before reaching the amplification stage (see \cite{MON08}).

\begin{figure}
\begin{center}
\includegraphics[width=12 cm]{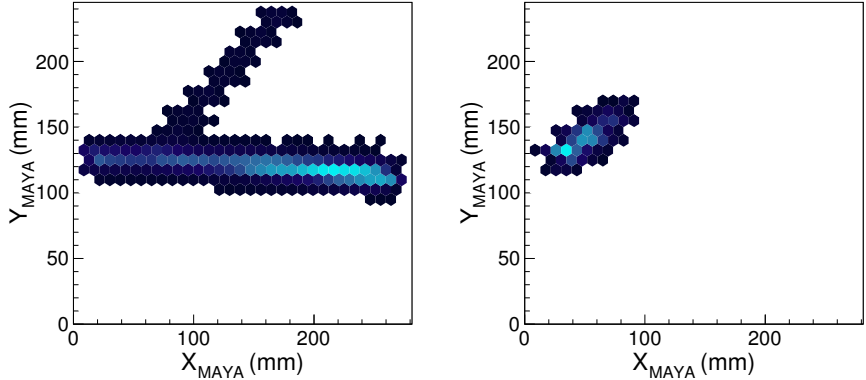}
\caption{\label{Fig.4} (Color online) Two-dimensional projection of the charge distribution measured from a $^{1}$H($^{11}$Li,$^{9}$Li)t reaction in \mbox{150 mbar} of isobutane (left panel) and from a $^{13}$N nucleus of \mbox{8.7 MeV} in \mbox{30 mbar} of isobutane produced in the $^{12}$C($^{8}$He,$^{13}$N)$^{7}$H reaction \cite{CAA06} (right panel)}
\end{center}
\end{figure}

- Multi-track setups refer to configurations where both beam and recoil particles contribute to the recorded pattern (as in \cite{TAN08}). Examples of charge distributions measured in such cases are presented in Fig. \ref{Fig.4}. 

\section{Trajectory Reconstruction}

In a first stage, the data analysis aims at extracting the direction of the trajectories from the induced patterns. However, no universal tracking algorithm can be used to reconstruct all the different types of pattern found in the experiments performed. Two basic methods, referred as the \textit{Hyperbolic Secant Squared} and the \textit{Global Fitting} methods, proved to be useful in most of the cases. They are reviewed in the following sections. 

\subsection{The Hyperbolic Secant Squared method}
\label{SECHS} 

\begin{figure}
\begin{center}
\includegraphics[width=12 cm]{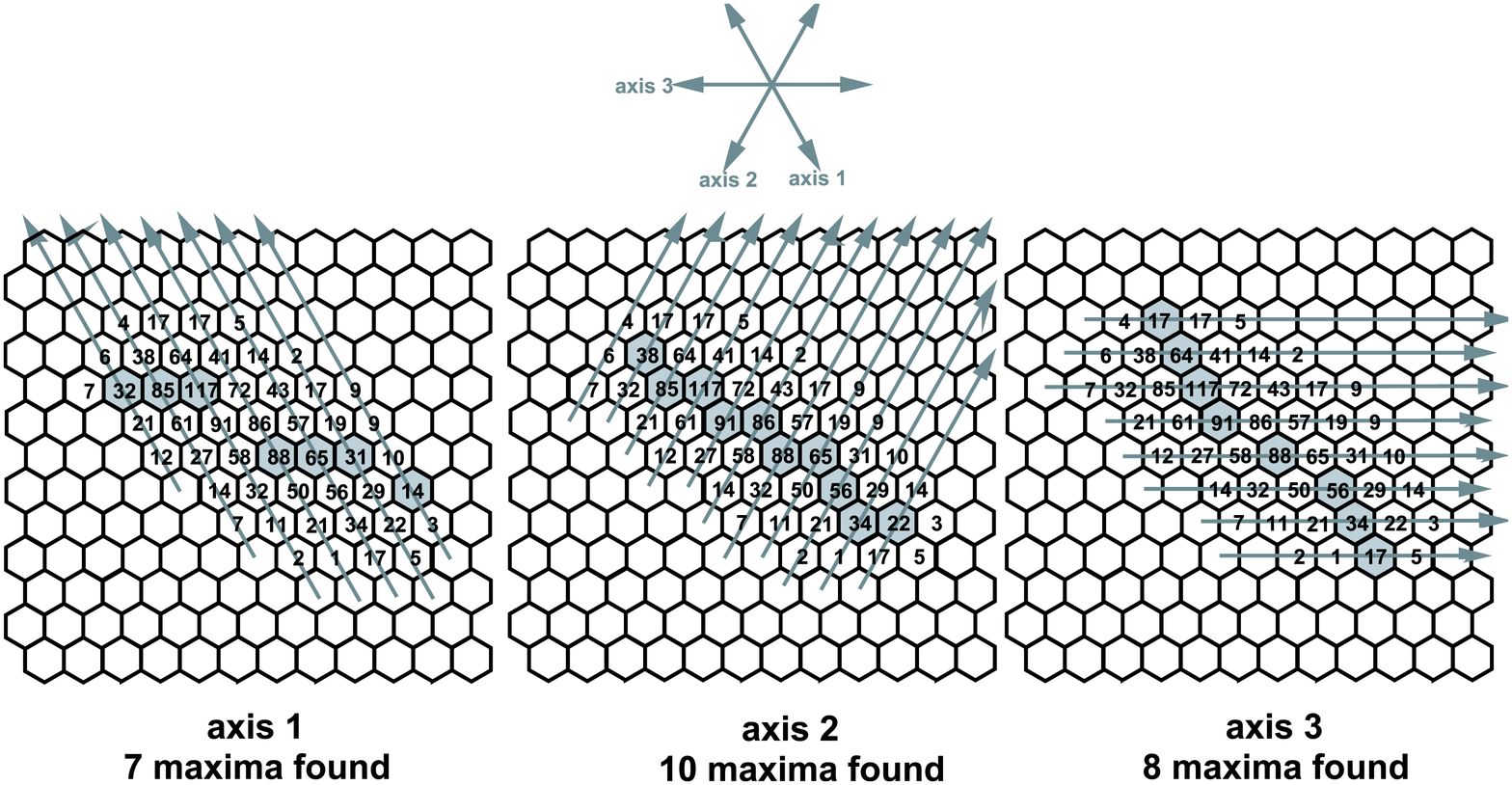}
\caption{\label{Fig.5} Illustration of the algorithm used to choose the optimum axis for the reconstruction of the track. In the picture, it corresponds to the axis labeled as ``axis 2'', which yields the highest number of maxima found.}
\end{center}
\end{figure}

The \textit{Hyperbolic Secant Squared} (SECHS) method is based on the determination and selection of the intersection points of the particle trajectory with the three symmetry axis of the cathode, defined by the hexagonal shape of the pads (see Fig. \ref{Fig.5}). The selected points are used to fit a straight line, which corresponds to the projected trajectory of the particles.

The first step is to identify the maxima of the collected charge, \textit{i.e.} the highest charge with two non-zero neighboring charges, along each symmetry axis. Figure \ref{Fig.5} shows the search of maxima over an experimental track measured in Ref. \cite{CAA07}. Once the pads with maximum charge are identified, the intersection point is estimated from the position of the pad and the centroid of the charge distributed between the pad and its two immediate neighbors. This is done using the SECHS formula \cite{LAU95} in a modified version \cite{DEM02}:

\begin{equation}
\Delta_{R}=\frac{w}{2}\frac{\ln\left(\frac{1+a_{1}}{1-a_{1}}\right)}{\ln\left(a_{2}+\sqrt{a_{2}^{2}-1}\right)}
\end{equation}
\begin{displaymath}
\mathrm{with~~~} a_{1}=\frac{\sqrt{\frac{Q_{0}}{Q_{+}}}-\sqrt{\frac{Q_{0}}{Q_{-}}}}{2\sinh a_{2}}\mathrm{~~~and~~~} a_{2}=\frac{1}{2}\left(\sqrt{\frac{Q_{0}}{Q_{+}}}+\sqrt{\frac{Q_{0}}{Q_{-}}}\right)
\end{displaymath}
where $\Delta_{R}$ is the distance between the estimated position of the centroid and the center of the pad with the maximum charge, $Q_{0}$. $Q_{+}$ and $Q_{-}$ are the charges measured on the left and right neighboring pads, and $w$ is the pad width. The particle trajectory, \textit{i.e.} its projected angle and position, are then fitted from the resulting centroids, separately for each symmetry axis. In order to avoid heavy data processing, the symmetry axis with the highest number of maxima was chosen to follow the whole method in some of the previous analysis (see Fig. \ref{Fig.5}). 

\begin{figure}
\begin{center}
\includegraphics[width=10cm]{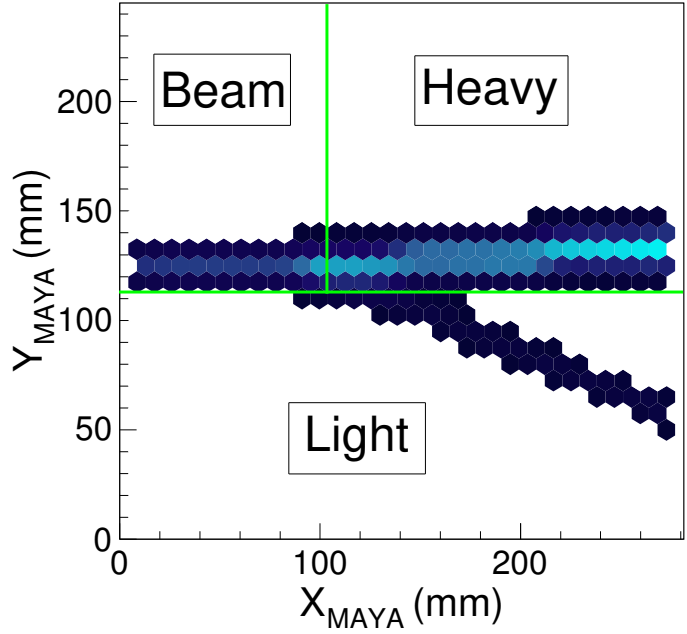}
\caption{\label{Fig.6} (Color online) Separation of the cathode plane in three different zones used to reconstruct the trajectories of the different charged particles.}
\end{center}
\end{figure}

Another adaption of this process consists in the definition of different areas of search when there are multiple particles involved in the reaction. For single-track setups, the search is performed over the whole cathode plane, whereas for multi-track setups different areas that encompass each individual track have to be considered. Usually, the beam track is separated from the reaction products dividing the cathode plane into three areas along the beam axis, as shown in Fig. \ref{Fig.6}. Other division patterns may be applied depending on the particles detected in the cathode.

\begin{figure}
\begin{center}
\includegraphics[width=6cm]{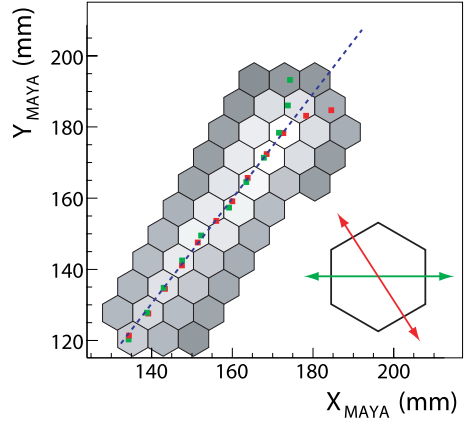}
\caption{\label{Fig.7} (Color online) The figure shows the reconstruction problem of the SECHS method when approaching the end of the track for an angle of \mbox{60 deg}. The dashed line represents the real trajectory, red and green squares are the calculated centroids for two different axis of symmetry.}
\end{center}
\end{figure}

\begin{figure}
\begin{center}
\includegraphics[width=12 cm]{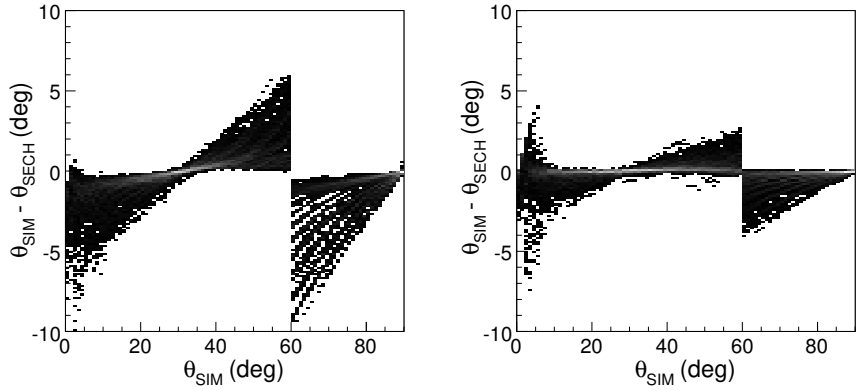}
\caption{\label{Fig.8} The absolute error on the reconstructed angles in the case of a single track setup is shown when using the SECHS method with all calculated centroids (left panel) and removing the two first and last centroids (right panel).}
\end{center}
\end{figure}

The angular resolution of this method depends strongly on the angle between the trajectory and the symmetry axis. The simulated resolution in single-track setups ranges between \mbox{0 deg} for trajectories perpendicular to any of the symmetry axis (\textit{i.e.} trajectories with 30, 90, or \mbox{150 deg} respect to the beam line), and \mbox{1 deg} for those parallel (\textit{i.e.} trajectories with 0, 60 or \mbox{120 deg}. The reason for this difference is illustrated in Fig. \ref{Fig.7}: when approaching the end of the track the reconstructed maxima of charge deviate from the trajectory, the charge distribution along the chosen symmetry axis being not described by the SECHS function anymore, producing shifted centroids, and introducing a systematic error in the fitting process. This effect can be reduced with a combined use of the centroids from more than one symmetry axis, this approach increases the resolution and reduces the error to less than 0.5 deg (see Fig. \ref{Fig.8}). Another possibility, if the track is long enough, is to exclude these points, which correspond approximately to the two first and two last centroids in the case of the single-track setups, and the two last ones for the multi-track setups. This operation reduces the uncertainty to less than \mbox{0.2 deg} (see Fig. \ref{Fig.8}). For multi-track setups, only the end of the track is available and therefore the effect of the shifting centroids is diminished. In these situations, the uncertainty is reduced approximately in a factor two, compared to the single-track setups.

\subsection{The Global Fitting method}
\label{FIT}

\begin{figure}
\begin{center}
\includegraphics[width=6 cm]{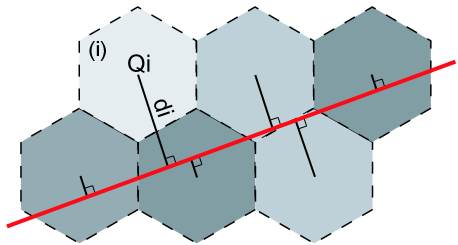}
\caption{\label{Fig.9} (Color online) Principle of the \textit{Global Fitting method}. The orthogonal distance of the center of the pads to the straight line, weighted by their charge is minimized.}
\end{center}
\end{figure}

The use of the SECHS method requires the presence of maxima in the charge pattern. However, this is not always the case. In reactions involving light particles with relatively high energy, the charge induced is not spread enough over the pads, due to the energy-loss profile of such particles. In this particular case, a different method is used, based on a fit of the whole charge distribution by means of the orthogonal distance regression procedure \cite{BOG88}. This method aims at finding the parameters of a first-degree polynomial minimizing the value of:
\begin{equation} 
\chi^{2}=\sum_{n=0}^{N}Q_{n}\frac{(a_{0}x_{n}+a_1-y_{n})^{2}}{a_0^{2}+1}
\end{equation}
where $a_0$ and $a_1$ are the offset and slope of the polynomial, $Q_{n}$ is the collected charge of the $n$ pad, and $x_{n},y_{n}$ are its center coordinates (see Fig. \ref{Fig.9}).

\begin{figure}
\begin{center}
\includegraphics[width=12 cm]{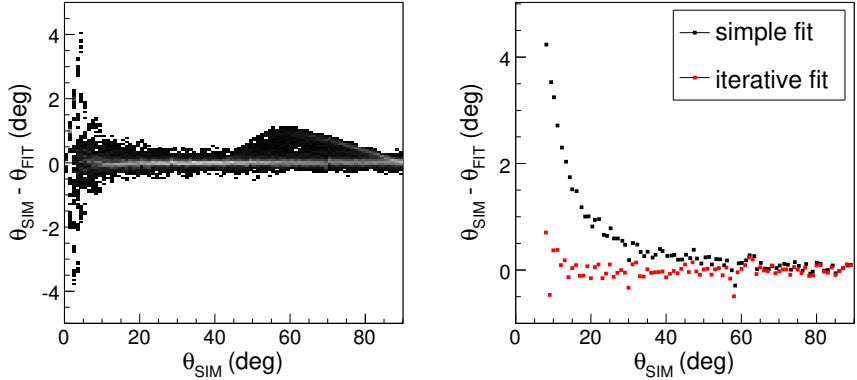}
\caption{\label{Fig.10} (Color online) The left panel depicts the error on the reconstructed angles using the FIT method in the case of single-track-setups. The right panel depicts the results of the FIT method when using a horizontal split of the cathode plane (black squares) and using an iterative cut (red squares, see text for details) in the case of multi-track setups.}
\end{center}
\end{figure}

\begin{figure}
\begin{center}
\includegraphics[width=12 cm]{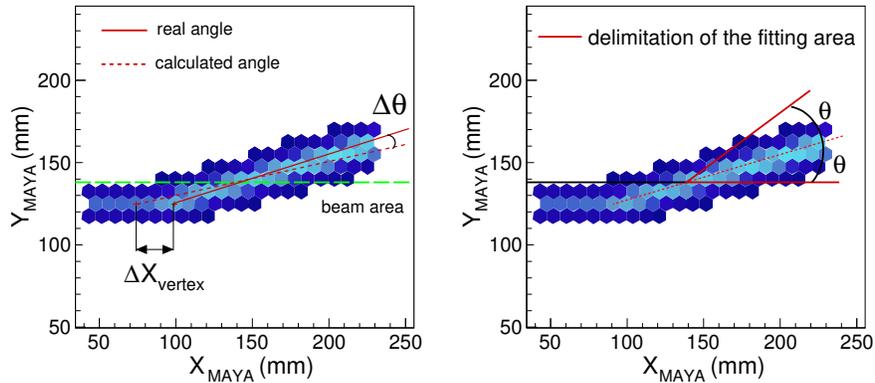}
\caption{\label{Fig.11} (Color online) The left figure shows a limitation of the FIT method in the case of a multi-track setup: the horizontal split at the beginning of the track makes the calculated angle being underestimated: only the upper side of the induction pattern contributes to the fit in the first part of the track. The right panel depicts the results of the FIT method when using a horizontal split of the cathode plane (black squares) and using an iterative cut (red squares, see text for details).}
\end{center}
\end{figure}

The resolution reached by this algorithm is better than 0.1 deg in single-track setups (see Fig. \ref{Fig.10}). For multi-track setups, the artificial splitting of the cathode plane in different regions for the different tracks leads to a decrease of the resolution. In the example illustrated in Fig. \ref{Fig.11}, the separation between regions allows only left pads to contribute at the beginning of the track. As a result, the calculated angle is underestimated. In general, the error on the calculated angle decreases with larger ranges and/or angles (see Fig. \ref{Fig.10}). In order to diminish this effect, an artificial cut in the side of the track opposite to the beam can be used to balance this effect. The cut is a straight line with an angle respect to the fitted trajectory equal to that between the trajectory and the beam line. A second fit is performed with the resulting charge pattern, and a second cut is applied in the same way as the first one with the new fitted angle. The process is repeated until the fitted angles converge (see Fig. \ref{Fig.11}). The result of this iterative process reduces the uncertainty to a maximum of $\sim$ 1 deg for angles around 10 deg (see Fig. \ref{Fig.10}). Below this value, the {\it Global Fitting method} is unsuitable.

\section{\label{Stopping}Range Measurement}

The total energy of particles stopping inside the filling gas of the detector is measured with the determination of the range. The link between both quantities, range and energy, depends on the mass and atomic charge of the particle, and the characteristics of the filling gas (composition and pressure). In order to measure the range it is necessary to determine the starting and stopping points. 

\begin{figure}
\begin{center}
\includegraphics[width=8 cm]{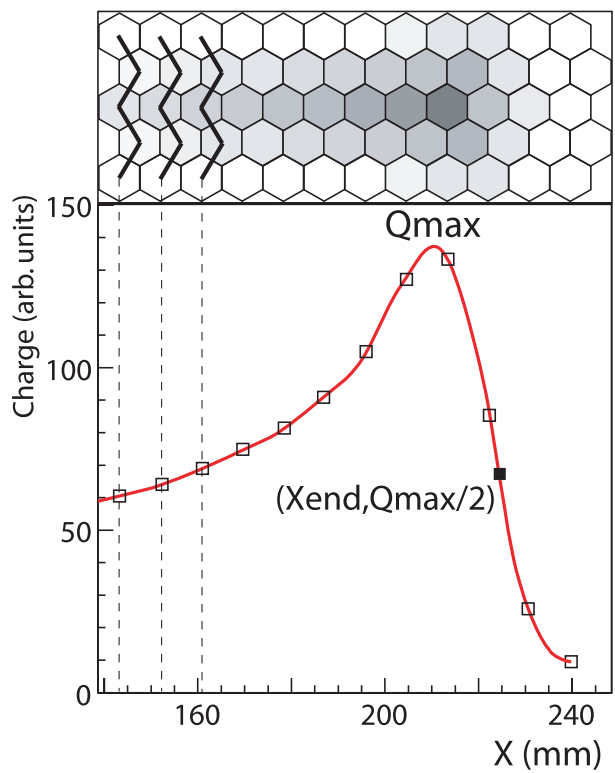}
\caption{\label{Fig.12} (Color online) Illustration of the technique used to build the charge profile and to reconstruct the ending point of the particles trajectories.}
\end{center}
\end{figure}

The measurement of the energy loss along the particle track, built by summing the charges collected on the pads along the axis more perpendicular to the trajectory or the pads belonging to same column in the case of 0 deg tracks, defines the \textit{charge profile} of the particle. The starting and stopping points are extracted from the analysis of the measured charge profiles. Figure \ref{Fig.12} shows an example for a particle at 0 deg, along the beam direction. Two different methods to extract the stopping point and other two to determine the starting point from the charge profile are explained in this section. 

\subsection{Stopping point determination}
\begin{figure}
\begin{center}
\includegraphics[width=12 cm]{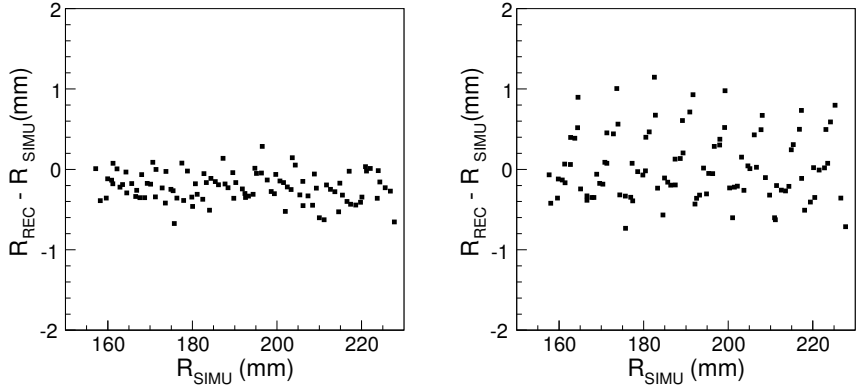}
\caption{\label{Fig.13} The panels show the deviation on the ending point reconstruction with the presented algorithm for different values of the threshold in the pads. A threshold on 5$\%$ of the Bragg peak height (left), shows no periodic effect on the reconstruction. For a value of 10$\%$ of the Bragg peak, a digitalization effect appears (right).}
\end{center}
\end{figure}

The first method to determine the stopping point focuses on the Bragg peak, a very well-defined point in the charge profile. The stopping point of the trajectory is defined as the point after the Bragg peak with an energy-loss half of the maximum charge of the profile. This point is an approximate of the real trajectory end point. Comparisons with TRIM calculations showed that the error made is, in average, less than 0.2 mm as shown on Fig. \ref{Fig.13}. A cubic spline interpolation is used to smooth the charge profile and extract the position of this point with the minimum error.

This procedure has been used in the measurement of the mass of $^{11}$Li by the Q-value of $^{1}$H($^{11}$Li,$^{9}$Li)t \cite{ROG08}, and tested on simulated patterns corresponding $^{11}$Li ions at 0 deg with different energies. The difference between the reconstructed and the actual range is kept well below 0.5 mm. However, the accuracy of this method is found to depend strongly on the detection threshold of the pads. The variation of the threshold has an immediate effect on the detection of the last point of the charge profile, and therefore on the spline behavior. If this charge is not detected the error may increase up to 1 mm, and reveals the pad structure producing a relatively strong digitalization with a period equal to the size of the pads (see Fig. \ref{Fig.13}).

A second method was developed for those cases where the Bragg peak is not formed in the charge profile because its width is of the order of the particle range. These situations appear, for example, with isotopes of medium to heavy mass with very low energy, of the order or lower than \mbox{1 MeV/u}. The short ranges involved do not allow for the use of a spline smoothing, and they force to extract the stopping point from the minimum number of charges. The charge collected in each step of the charge profile for a fixed range and energy depends on the position relative to the pads, and on the angles respect to the pad plane. Therefore, the relative charges are related with the stopping point position. Different combinations of the charges collected can be used to determine the stopping point. As an example, the stopping point can be estimated from a derivative of the charge profile, defined in the form of center of gravity of charge variations with the three last charges:

\begin{equation} 
Pos_{stop}=\frac{Pos_{last}+Pos_{last-1}}{2}+\Delta\left( d_{stop}\times CoG_{stop} + Off_{stop}\right)
\end{equation}
\begin{displaymath}
{\rm with}~~~CoG_{stop}=\frac{\delta q_{last-1}- \delta q_{last-2}}{\delta q_{last-1}+\delta q_{last-2}}~~~{\rm and}~~~\delta q_{i}=q_{i+1}-q_{i}
\end{displaymath} 

In this case, the position of the stopping point ($Pos_{stop}$) is calculated as a correction of the position of the last charges ($Pos_{last}$ and $Pos_{last-1}$). This correction is a product of the effective width of the pads along the trajectory ($\Delta$) and the center of gravity ($CoG_{stop}$) of the last ($\delta q_{last}$) and the second-to-last ($\delta q_{last-1}$) variations of charge in the charge profile. Two parameters, $Off_{stop}$ and $d_{stop}$, include the dependency on the characteristics of the detector (pad size, distance from wires to the pad plane), and must be specifically determined for each experimental setup.

\begin{figure}
\begin{center}
\includegraphics[width=12 cm]{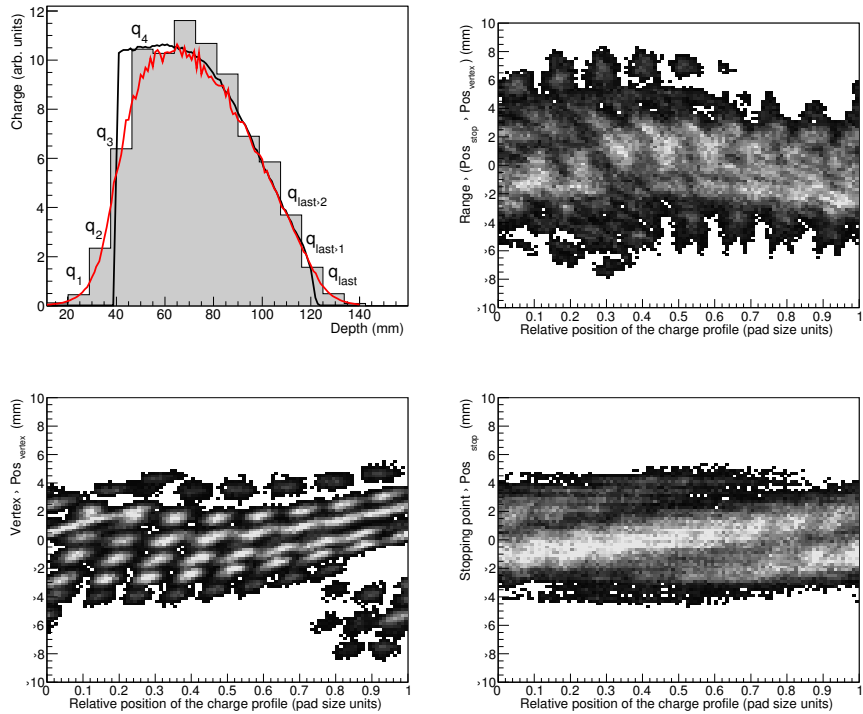}
\caption{\label{Fig.14} (Color online) The upper left figure shows a typical charge profile is displayed in the figure. The black line follows the energy-loss profile of the particle. The convolution of the energy-loss with the induction in the wires produces the profile in red. The full histogram is the resulting charge profile. The upper right and lower panels show the deviation of the range, the position of the vertex (bottom left), and the stopping point (bottom right), determined with the formulas described in the text as a function of the relative positions of the charge profile within the pads.}
\end{center}
\end{figure}

This formula was tested with simulated profiles of $^{13}$N of \mbox{0.5 MeV/u} in \mbox{30 mbar} of C$_{4}$H$_{10}$, with a pad size of 5 mm side and a distance of 10 mm between the wires and the cathode plane. A set of events was produced varying both the angle respect to the beam direction and the angle of the reaction plane. Figure \ref{Fig.14} shows an example of one of these profiles. The parameters $Off_{stop}$ and $d_{stop}$ were found to be $0.9$ and $1.5$. The resolution results of the order of $\sim$ 1.7 mm. Figure \ref{Fig.14} shows the difference between the simulated position of the stopping point and that determined with the previous formula versus the relative position of the trajectory within the pads. A characteristic pattern appears due to digitalization produced in the formation of the charge profile. The resolution of this method can be further improved by finding the parameters $Off_{stop}$ and $d_{stop}$ as functions of the angles respect to the beam direction and the angle of the reaction plane. 

An alternative to this procedure is to fit the expected charge profile, calculated with TRIM or similar codes and folded with the wires induction and geometry, to each of the measured ones. The heavy data processing involved prevented its use in the past.

\subsection{\label{Vertex}Reaction point determination}

The reaction point is an important observable as it gives information about the precise energy of reaction, accounting for the energy-loss of the projectile. In addition, it also corresponds to the starting point of the trajectories, which is necessary to measure the ranges of the detected reaction products.

In the case of multi-track setups, the reaction point corresponds to the intersection between the trajectories of, at least, two of the particles involved in the reaction. However, as shown in Fig. \ref{Fig.11}, when the angles respect to the beam axis of the particles are small (typically lower than \mbox{10 deg}) or when the range of the particle is small (typically less than 5 cm), the uncertainty from the angle reconstruction has a strong influence on the resolution of the vertex position. In these cases, the charge profile built along the central row of pads can be used to determine the vertex point. When the energy-loss of the products is different from that of the beam particles, the charge profile shows a sudden change at the reaction point. The shape of this change is related with the charge spread, and it also depends on the angle of the trajectories involved.

\begin{figure}
\begin{center}
\includegraphics[width=12 cm]{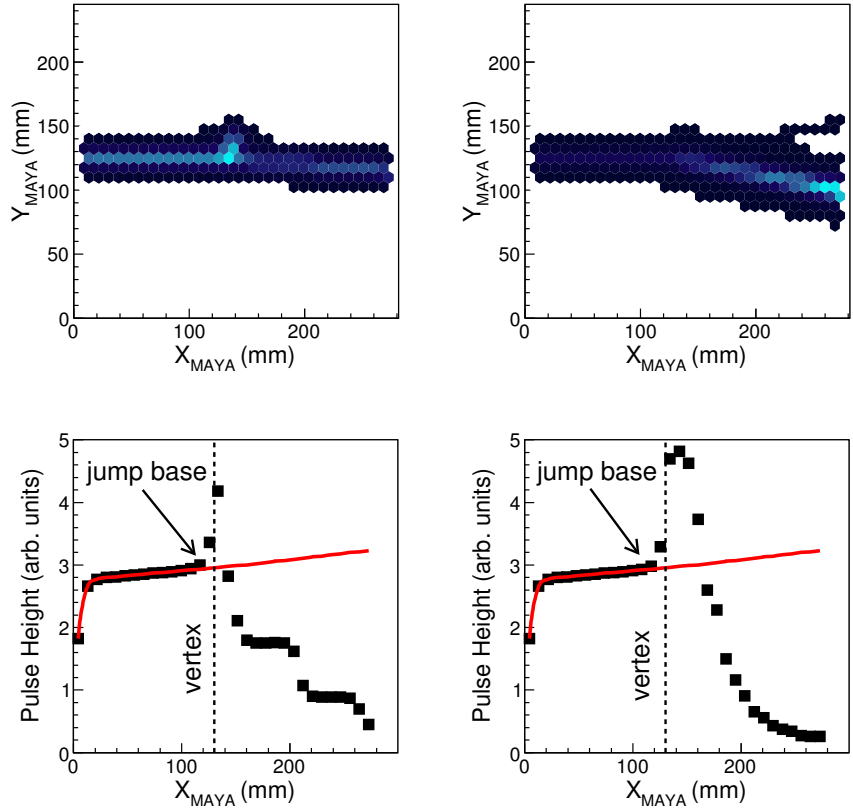}
\caption{\label{Fig.15} (Color online) Illustration of the technique for measuring the vertex position with short ranges and/or small angles. The picture shows the vertex on a $^{1}$H($^{11}$Li,$^{9}$Li reaction at 4.3\textit{A} MeV at 10$^{\circ}$ (upper panels) and 150$^{\circ}$ in the center of mass (lower panels). Both reactions take place at 100 mm from the entrance of the detector. The observed charge profile is represented by the black squares, while the expected charge profile without any reaction is represented by the red curve. The vertex point is found at twice the pad size from the base of the jump in the charge profile. This distance corresponds to the width of the charge spread in the induction process.}
\end{center}
\end{figure}

\begin{figure}
\begin{center}
\includegraphics[width=6 cm]{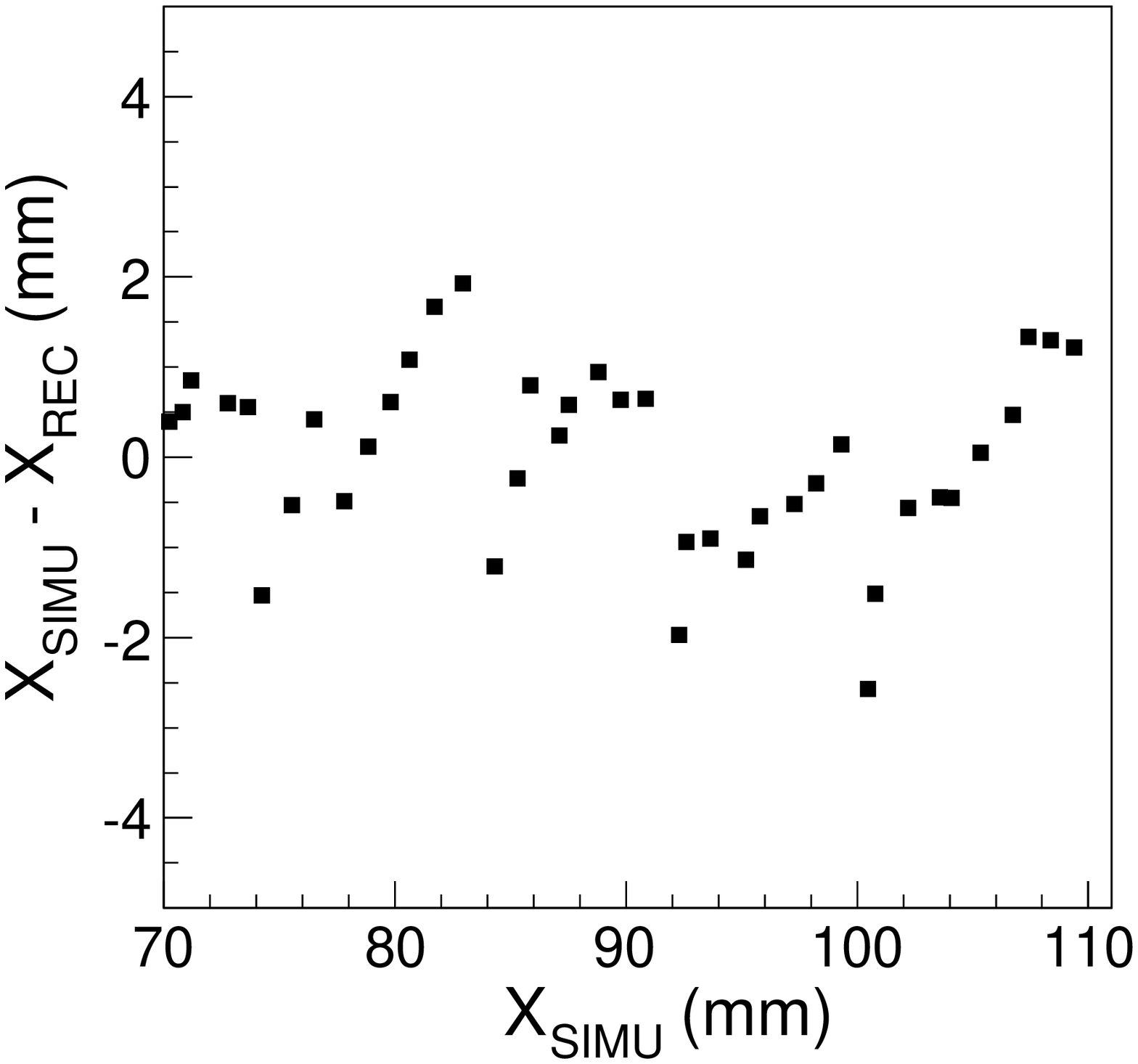}
\caption{\label{Fig.16} The figure shows the simulated uncertainty in the determination of the vertex position with the charge profile for a set of multi-track setups events from $^{1}$H($^{11}$Li,$^{9}$Li). See text for details.}
\end{center}
\end{figure}

Figure \ref{Fig.15} shows an example of a transfer reaction in inverse kinematics at extreme center-of-mass angles, $^{1}$H($^{11}$Li,$^{9}$Li) at \mbox{4.3\textit{A} MeV} for \mbox{20 deg} and \mbox{150 deg} in center-of-mass frame, where this method is well-suited. A sudden increase in the charge profile appears near the vertex point, due to the different energy-loss of the beam and the reaction products. Considering $W$ as the width of the charge spread, the vertex point is then defined as the point located at a distance $W$ from the base of the sudden increase of the collected charge. The maximum error with this method, for the $^{1}$H($^{11}$Li,$^{9}$Li) reaction, is around 2 mm (see Fig. \ref{Fig.16}). In general, the performance of this method depends on the characteristics of the reaction and detected particles involved.

Single-track setups do not detect beam particles, therefore the reaction point must be deduced with the charge profile of the single detected particle. Here, a similar technique to that used for the stopping point determination can be applied. The vertex position can be estimated from a discreet derivative of the charge profile, defined as the variation of the two first charges ($\delta q_{1}$), normalized to the variations in the three first charges ($\delta q_{1}+\delta q_{2}$):

\begin{equation} 
Pos_{vertex}=Pos_{4}+\Delta\left( d_{vertex}\times CoG_{vertex} + Off_{vertex}\right)
\end{equation}
\begin{displaymath}
{\rm with}~~~CoG_{vertex}=\frac{\delta q_{1}}{\delta q_{1}+\delta q_{2}}~~~{\rm and}~~~\delta q_{i}=q_{i+1}-q_{i}
\end{displaymath}

The position of the vertex ($Pos_{vertex}$) is then calculated as a correction of the position of the fourth charge ($Pos_{4}$). This correction is a product of the width of the pads along the trajectory ($\Delta$) and the normalized variation of the first charges ($CoG_{vertex}$) of the charge profile. Two parameters, $Off_{vertex}$ and $d_{vertex}$, include the dependency on the characteristics of the detector, and must be again specifically determined for each experimental setup.

This formula was also tested with simulated profiles of $^{13}$N of \mbox{0.5 MeV/u} in \mbox{30 mbar} of C$_{4}$H$_{10}$, with a pad size of 5 mm side and a distance of 10 mm between the wires and the cathode plane. The parameters $Off_{vertex}$ and $d_{vertex}$ were found to be $0.9$ and $1.5$. The resolution is of the order of 2.0 mm (see Fig. \ref{Fig.14}). Again, this resolution can be reduced to $\sim$1.5 mm if the parameters $Off_{vertex}$ and $d_{vertex}$ are free to vary with the angles of the trajectory. 

The overall resolution on the determination of the range is around the quadratic sum of the uncertainties in both the vertex and the stopping point, which in the case of the simulated events, corresponds to $\sim$2.6 mm (see Fig. \ref{Fig.14}).

\section*{Conclusions}

Different methods for tracking reconstruction for two-dimentionnal projected tracks, already applied to data measured with the active target detector MAYA, are reported. The procedures determine projected angles, as well as the stopping and starting points of the measured tracks. The resolution of these measurements are extracted from simulations of realistic experimental setups, covering the different experiments performed with MAYA. These reconstruction techniques may be used in similar detectors were tracking is done from segmented images of the trajectories.

\section*{Acknowledgements}

M. Caama\~no thanks the support of the spanish Centro Nacional de Física de Partículas, Astropartículas y Nuclear (CPAN)

\end{document}